\documentclass[a4paper,12pt]{article}

\usepackage[usenames,dvipsnames]{pstricks}
\usepackage{epsfig}

\usepackage[T2A]{fontenc}
\usepackage[utf8]{inputenc}   % older versions of ucs package
\usepackage[russian,english]{babel}

\usepackage{verbatim} 	% to be able to use \begin{comment} ... \end{comment}
\usepackage{amsmath, amsthm}
\usepackage{graphicx}
\usepackage{array} 	% needed for \setlength{\extrarowheight}{0.3cm} 
\usepackage{rotating}	% This allows you to rotate text, tables and figures in every manner.

\usepackage{hyperref}	% for hreferences. Use like this in document:
\begin{document}
\pagestyle{plain}
\pagenumbering{arabic}

\selectlanguage{english}

\title{Quantum-well states and discontinuities in opto-electrical characteristics of SCH lasers.}

\author{Z. Koziol\footnote{Corresponding author email: zbigniew@ostu.ru},\\
Orel State Technical University,\\
29 Naugorskoye Shosse, Orel, 302020, Russia.}

\maketitle

\begin{abstract}

Computer simulations with Synopsys' Sentaurus TCAD of opto-electrical characteristics 
of separate-confinement heterostructure laser based on $AlGaAs$ are used as an example 
to study the role of the width and depth of Quantum Well (QW) 
active region on laser characteristics, I-V and I-L, below and above the lasing threshold. 
The device properties depend on both, the number of bound QW states and on closeness 
of the highest bound states to conduction or valence band offset.
The lasing action may not exist at certain widths or hights of QW, 
the threshold current is a discontinuous function of these parameters.
The effects are more pronounced at low temperatures. Discontinuities in characteristics
may be observed at certain conditions in temperature dependencies of laser parameters.
\end{abstract}

%%%%%%%%%%%%%%%%%%%%%%%%%%%%%%%%%%%%%%%%%%%%%%%%%%%%%%%%%%%%%%%%%%%%%%%%%%%%%%%%%%%%%%%

\section{Introduction}

Computer modelling of electronic devices is a relatively 
new approach towards study of physical phenomena occurring there as well
optimizing their technical characteristics. Methodologicaly, this field of
scientific and engineering activity may be placed between theory and experiment:
persuing research of that kind requires theoretical understanding of physics
of microscopic processes and may work as a helpfull tool in interptretation 
of experimental data. However, contrary to sometime met thinking, computer modelling
can not replace either theory or experiment. But in some situations, results
of that research may provide an inspiration for understanding or testing physical phenomena: it is
easier, faster and less expensive to perform modelling than experiments, and
we are not restricted that much by, often large, inaccuracy of experimental data that may hide
insightfull details. 

When performing modelling $AlGaAs$ SCH lasers with Synopsys' Sentaurus TCAD \cite{tcad}, we noticed unexpected steps 
in some of their characteristics (threshold current $I_{th}$ versus the width of quantum well $d_a$,
or versus its height, etc). Analyses of results led us to an idea that observed discontinuities 
occur when the most upper bound QW state crosses the conduction 
or valance band offset energy. Following that idea, we guessed that the effects may manifest
itself in temperature dependence of some quantities as well, if laser parameters are choosen for that
properly. The discontinuities are found also below the lasing threshold current,
in their $I-V$ characteristics or in gain (or loss) versus current. 

We attribute the existence of observed effects to tunneling transport phenomena through QW.
When the dimensions of the structure are reduced below the mean free path 
(which is of the order of $10^2$ - $10^4$ $nm$ for AlGaAs), one enters
the ballistic transport regime; drift-diffusion model of carriers transport must be replaced 
with transfer method type of calculations when dealing with QW widths 
of the order of $10 nm$ \cite{Beenakker}.

%%%%%%%%%%%%%%%%%%%%%%%%%%%%%%%%%%%%%%%%%%%%%%%%%%%%%%%%%%%%%%%%%%%%%%%%%%%%%%%%%%%%%%%%%%%%%%%
\section{Modelling}

\begin{table}[t]
\caption{A few sets of simulation conditions ($A-F$) 
for data shown in Figures \ref{qw300_talk0A} and \ref{qw300_talk1}. 
$C$ is radiative recombination rate (Eq. \ref{Radiative_recombination_rate}).
$\alpha_n$, and $\alpha_p$ are coefficients of free carrier absorption formula (\ref{Free_carrier_absorption}).
Temperature for all cases is $T=300 K$, electron and hole scattering times are assumed 
$8\cdot 10^{-13} s$ and $4\cdot 10^{-13} s$, respectively, and electron and hole mobility 
$9200 cm^2/Vs$ and $400 cm^2/Vs$, respectively. No $SRH$ recombination and no additional 
light scattering mechanisms are assumed.
}
\begin{center}
\begin{tabular}{|c|c|c|c|c|c|}
\hline
\#  		&	C [$cm^3/s$] 	&  $\alpha_n$ [$cm^{-2}$]	&  $\alpha_p$ [$cm^{-2}$]\\
\hline
A 			&	$2.0 \cdot 10^{-10}$	&	$1 \cdot 10^{-18}$		&	$2 \cdot 10^{-18}$	\\
\hline
B 			&	$2.0 \cdot 10^{-10}$	&	$5 \cdot 10^{-17}$		&	$1 \cdot 10^{-18}$	\\
\hline
C 			&	$2.0 \cdot 10^{-10}$	&	$1.5 \cdot 10^{-18}$		&	$3 \cdot 10^{-18}$	\\
\hline
D 			&	$1.0 \cdot 10^{-10}$	&	$1.5 \cdot 10^{-18}$		&	$3 \cdot 10^{-18}$	\\
\hline
E 			&	0 							&	$1.5 \cdot 10^{-18}$		&	$3 \cdot 10^{-18}$	\\
\hline
F 			&	0							&	$1.5 \cdot 10^{-18}$		&	$3 \cdot 10^{-18}$	\\
\hline
\end{tabular}
\end{center}
\label{table_4}
\end{table}

The laser we are modelling has dimensions, structure and doping 
as described by Andreev, et al. \cite{Andrejev_1}, \cite{Andrejev_2}.
The lasing wavelength is $808 nm$. The lasing offset voltage $U_0$ is $1.56-1.60 V$, 
differential resistance just above the lasing offset, $r=dU/dI$ is $50-80 m\Omega$, threshold current $I_{th}$ is $200-300 mA$,
slope of optical power, $S=dL/dI$ is $1.15-1.25 W/A$, and left and right mirror reflection 
coefficients $R_l$ and $R_r$ are $0.05$ and $0.95$. The reference laser has the width of QW, $d_a$, 
of $12 nm$ and both waveguides' width is $0.2 \mu m$. 
In order to reproduce laser characteristics in computational results, we
played with several variables available in Synopsys. The critical one is $A_{ph}$ 
- the effective surface area factor in $Physics$ section of Synopsys command file. 
An agreement between experiment and calculation is reached for $A_{ph}$ of about $0.059$. 
This low value of $A_{ph}$ should not be surprising. If carrier drift-diffusion processes only
were responsible for transport than $A_{ph}$ close to $1$ would be expected. However, 
we are dealing here, in particular in waveguide and in QW regions, with ballistic transport as well, 
while the computational model used in Synopsys is derived from drift-diffusion equations, 
modified for dealing with transport through QW as discussed in the next section.

Other parameters available in Synopsys, important in this case, are these related 
to light absorption and carrier scattering. Experiment shows that absorption coefficient 
is of the order of $1 cm^{-1}$ (\cite{Andrejev_1}). It is argued that in AlGaAs lasers 
the main contribution to absorption is due to scattering on free carriers. 
The free carrier absorption coefficient, $\alpha_{fc}$, is given by:

\begin{equation}\label{Free_carrier_absorption}
	\alpha_{fc} = \left(\alpha_n \cdot n + \alpha_p \cdot p\right) \cdot L,
\end{equation}

where $n$ and $p$ are the electron and hole density, and $L$ is light intensity. We choosed in our calculations such values of $\alpha_n$ and 
$\alpha_p$ that an effective absorption coefficient would be obtained close to that experimental one.

It is important also to have a reasonable value of radiative recombination rate, 
$R_r$, which is assumed to be described by: 

\begin{equation}\label{Radiative_recombination_rate}
	R_{r} = C \cdot \left(n \cdot p - n_{i_{eff}}^2\right),
\end{equation}

where $n_{i_{eff}}$ describe the effective intrinsic density, and $C$ is a parameter available for changes.

\begin{figure}[t]
\begin{center}
      \resizebox{150mm}{!}{\includegraphics{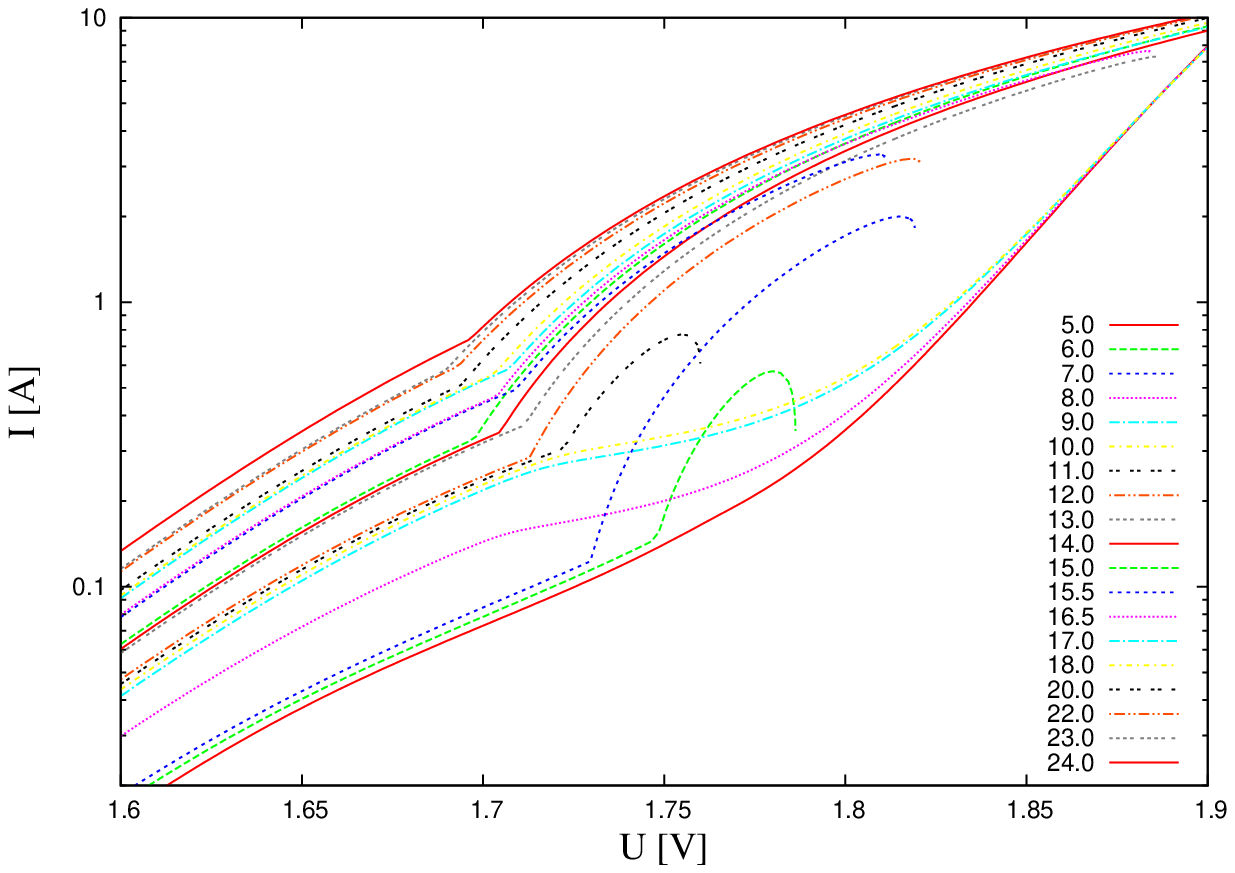}}
      \caption{
	Typical $I-V$ characteristics computed at $T=300$. $OpticalLoss$ parameter is assumed $0$,
	no radiative recombination, free carrier scattering rate parameters $\tau_n$ and $\tau_p$
	are $8 \cdot 10^{-13} s^{-1}$and  $4\cdot 10^{-13}  s^{-1}$, with electron and hole mobolities
	$9200 cm^2/Vs$ and $400 cm^2/Vs$. 
	The legend describes width of QW (in $nm$), from $5 nm$ from right-bottom curve to 
	$24 nm$ for uppermost curve.
}
	 \label{qw300_000}
\end{center}
\end{figure}

Typical $I-V$ characteristics computed at $T=300$ are shown in Figure \ref{qw300_000}, 
for a broad range of QW widths. For current 
near the lasing threshold current $I_{th}$ (i.e. for voltage near the lasing offset voltage $U_0$), 
which correspond to a kink in $I-V$, for most of these curves 
the results are very well approximated by a phenomenological modified exponential relation (\cite{Koziol}):

\begin{equation}\label{exponential_6_parameters}
\begin{array}{ll}
        I(U) = I_{th} \cdot exp(A\cdot (U-U_0) + B \cdot (U-U_0)^2),~~~ for ~ U< U_0\\
        I(U) = I_{th} \cdot exp(C\cdot (U-U_0) + D \cdot (U-U_0)^2),~~~ for ~ U> U_0
\end{array}
\end{equation}

where $I_{th}$, $U_0$, as well $A$, $B$, $C$, and $D$ are certain fiting parameters. 

There is no simple, 
straightforward interpretation of these curves above the lasing threshold, where a strong interplay between 
the effects of carrier transport and scattering takes place, with light absorption as well. However, 
we may notice an interesting feature for parts of curves below the lasing threshold. While
the width of QW, $d_a$, changes (nearly) monotonically, the curves however are grouped 
into a few sets such that they nearly coincide together within each group.

\begin{figure}[t]\begin{center}
      \resizebox{140mm}{!}{\includegraphics{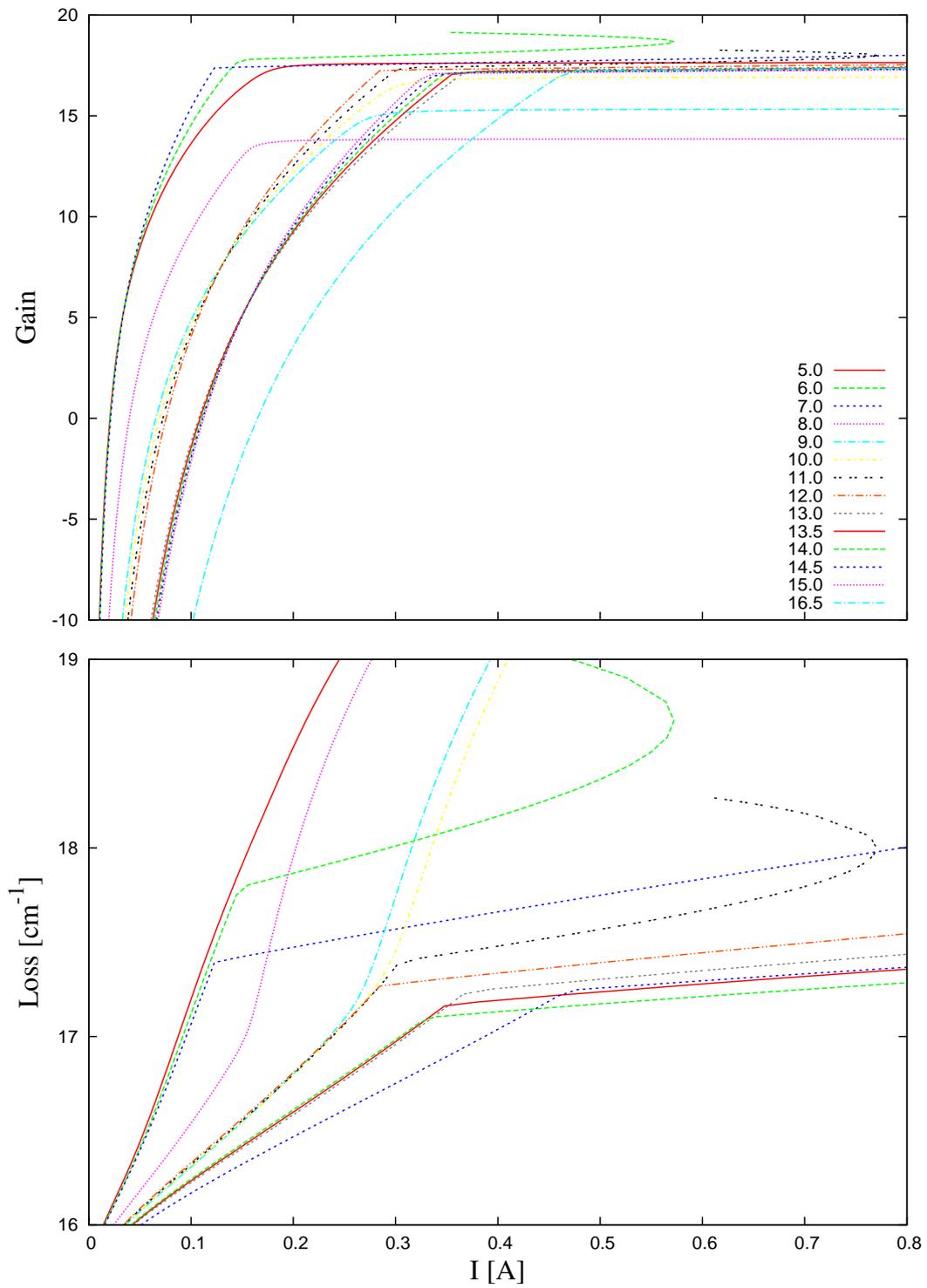}}
      \caption{
	Gain (upper figure) and loss computed for in similar conditions as the data in Figure \ref{qw300_000}.
	The legend describes width of QW (in $nm$).
}
\label{qw300_014A}
\end{center}\end{figure}

A very similar feature is observed when gain or loss is drawn as a function of current, for many
widths of active region (Figure \ref{qw300_014A}). Again, we see that datacurves for current
below the lasing threshold (that corresponds to kinks in curves) 
are grouped into a few sets such that they nearly coincide together within each group.
If gain or loss were drawn as a function of voltage, however, we would not see such a grouping.

Therefore, we conclude that below the lasing threshold, 
current as a function of QW width at constant voltage derived from the data like these in Fig. \ref{qw300_000},
or gain or loss as a function of QW width, at constant current values, also below the lasing threshold,
will follow step-like functions.

\begin{figure}[t]\begin{center}
      \resizebox{150mm}{!}{\includegraphics{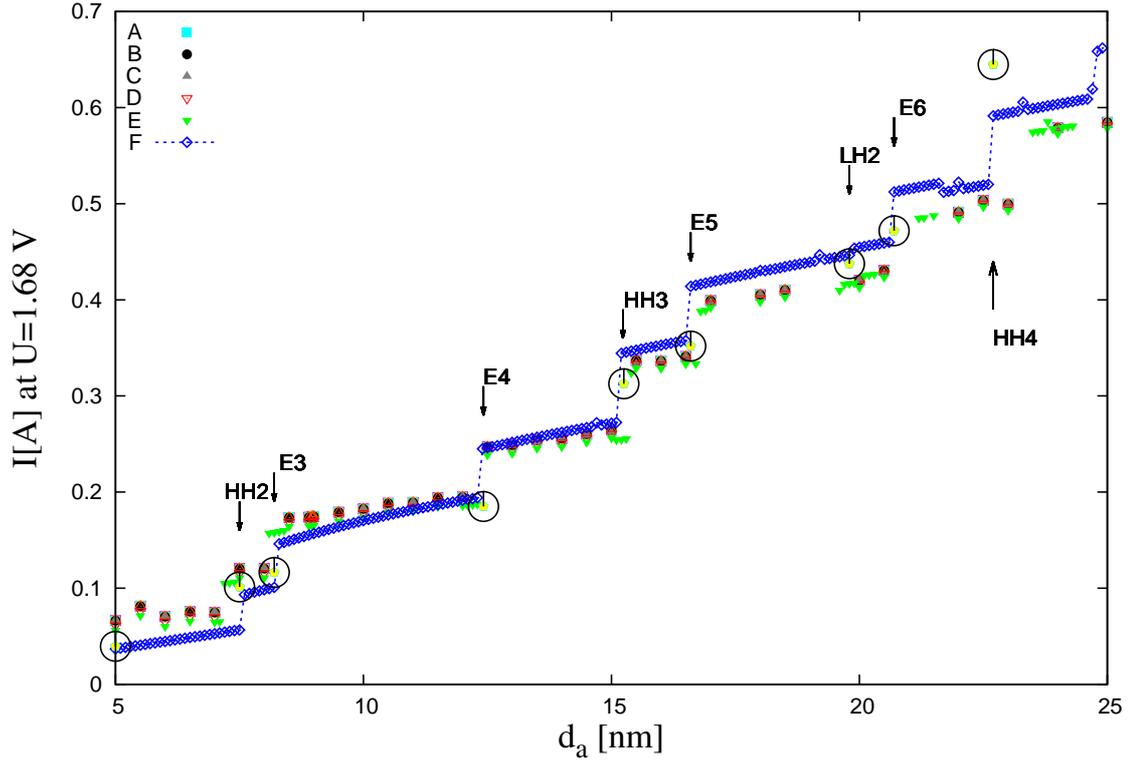}}
      \caption{
	Current as a function of QW width derived at constant voltage from datacurves 
	as these shown in Figure \ref{qw300_000} and described in Table \ref{table_4}. 
	The solid line for datapoints $F$ is to guide the eyes, only. $F$ is computed for constant 
	$Al$ concentration in QW of $8\%$, while all other datasets ($A-E$) are computed with
	such a concentration of $Al$ in QW that lasing wavelength will remain constant ($808 nm$)
	when QW width changes. The arrows are for datapoints $F$.
}
\label{qw300_talk0A}
\end{center}\end{figure}

\begin{figure}[t]\begin{center}
      \resizebox{150mm}{!}{\includegraphics{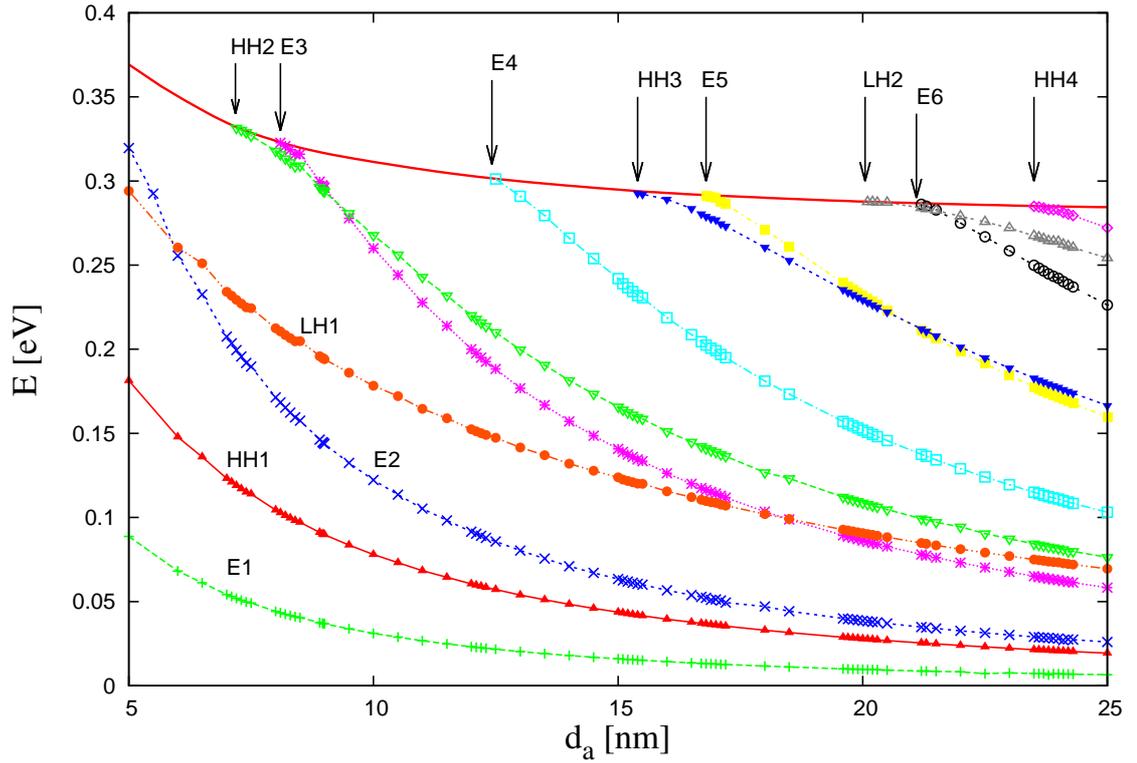}}
      \caption{
T=300K. Conduction and valence band offset energy, ($E_{CBO}$ and $E_{VBO}$), 
and electron ($E_n$), light- and heavy hole energies ($LH_n$ and $HH_n$) in QW, 
as a function of QW width. Hole energies and $E_{VBO}$ have been scaled up 
by a factor 28 to obtain coincidnce with electron energy scale ($E_{CBO}$ and $E_{VBO}$
curve are the same in this Figure).
}\label{qw300_energy0}
\end{center}\end{figure}

This is illustrated in Fig. \ref{qw300_talk0A}, where current as a function of QW width derived 
at constant voltage from datacurves similar to these as in Figure \ref{qw300_000} is shown. 
Figure \ref{qw300_talk0A} presents data computed at different conditions, and marked from $A$ to $F$, 
for several combinations of free carrier scattering coefficients, $\alpha_n$ and $\alpha_p$,
and values of $C$, the radiative recombination parameter, as described in Table \ref{table_4}.
Moreover, dataset $F$ differs from datasets $A$-$E$. The last are computed assuming changing 
Al concentration in waveguides (when Al in QW is kept constant) in such a way that the lasing 
wavelength does not change with the change of QW width. The dataset $F$ is computed for constant 
Al concentration in waveguides of $33 \%$. The solid line in Fig. \ref{qw300_talk0A} is drawn through
datapoints $F$, and arrows there refer to curve $F$ as well, and indicate positions of bound QW energy states crossing
the conduction- or valance band offset energies. Positions of bound QW energy states for cases $A$-$E$,
as illustrated in Fig. \ref{qw300_energy0}, are very close to but not identical.

\begin{figure}[t]\begin{center}
      \resizebox{150mm}{!}{\includegraphics{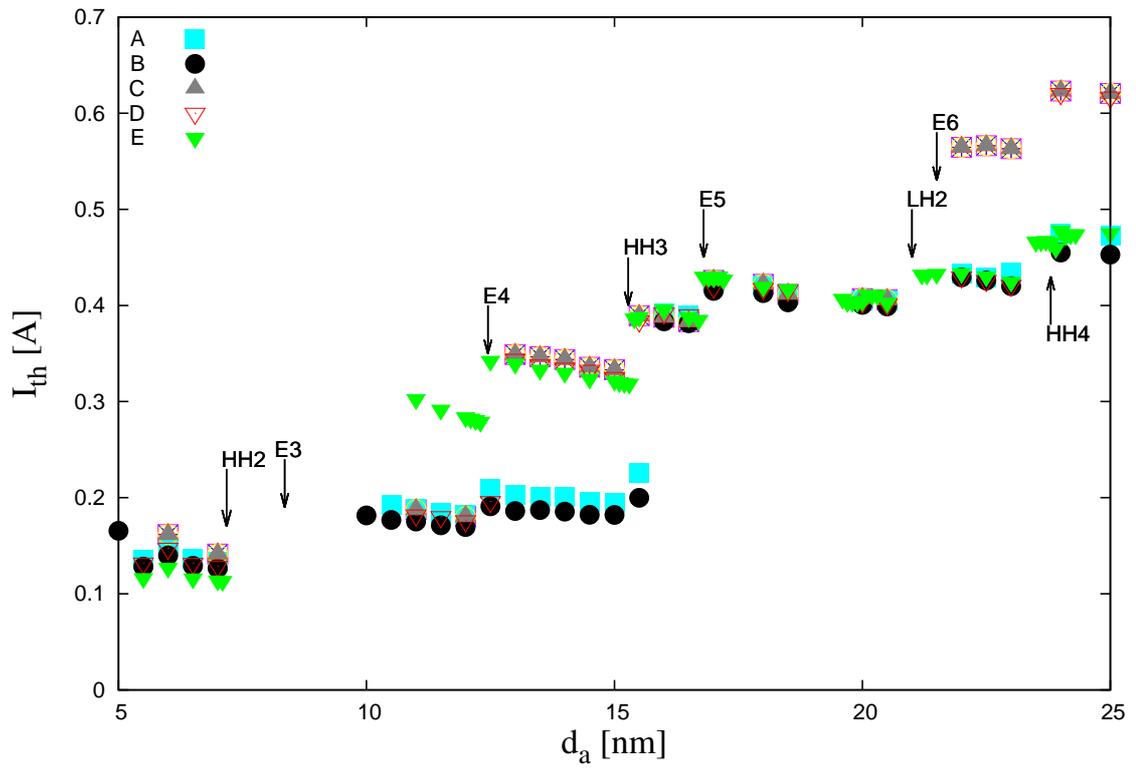}}
      \caption{
	Lasing threshold current as a function of QW width for datasets 
	as these shown in Figure \ref{qw300_000} and described in Table \ref{table_4}. 
	The arrows are at positions close to but not identical to these in Figure \ref{qw300_talk0A}.
}
\label{qw300_talk1}
\end{center}\end{figure}

The step-like features are preserved also in $I_{th}(d_a)$ dependencies, 
as illustrated in Fig. \ref{qw300_talk1}. There, however, effects of carrier scattering and light 
absorption smear-out the picture. It is useful to notice that at some values of QW width no lasing 
action is reached, and therefore the datapoints in that Figure are not available for all QW widths.

Changes of QW height (caused by difference of Al concentration in QW and waveguides) cause very 
similar step-like dependencies. Moreover, the effects are in some situations more clear and pronounced
at low temperatures. We did modelling for T=77.6 K to confirm their existence.

\begin{figure}[t]\begin{center}
      \resizebox{150mm}{!}{\includegraphics{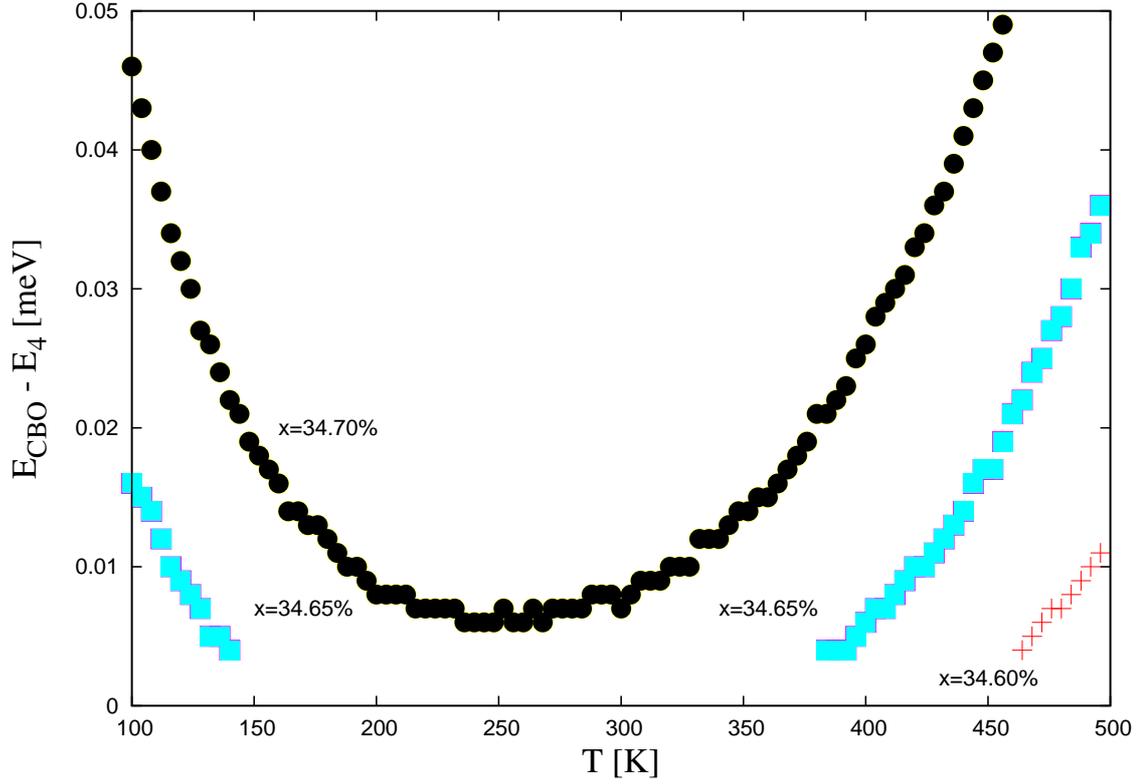}}
      \caption{The energy difference between conduction band offset energy $E_{CBO}$
and the closest to it bound electron state in QW, $E_4$, as a function of temperature, 
when active region concentration is 8\% of Al, for three values of waveguide Al concentraion:
$34.60\%$, $34.65 \%$, and $34.70\%$.
}
\label{temp-swap00}
\end{center}\end{figure}

\begin{figure}[t]\begin{center}
      \resizebox{150mm}{!}{\includegraphics{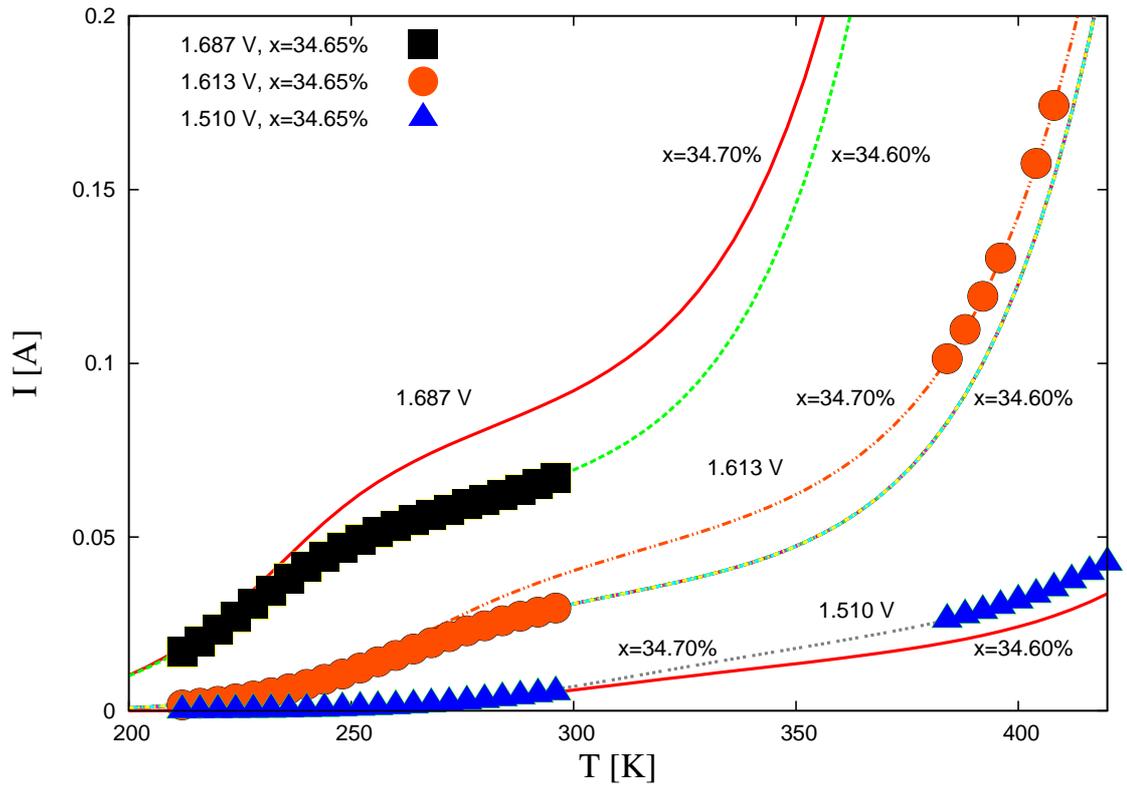}}
      \caption{Temperature dependence of current for free values of voltage applied 
(1.510V, 1.613V, and 1.687V; for each of these this is below the lasing threshold),
and for free values of $Al$ concentrations in waveguide, the same as in Figure \ref{temp-swap00}.
}
\label{temp-swap06}
\end{center}\end{figure}

Moreover, with a careful design of laser structure (content of Al in QW and waveguides) it is possible to 
find the evidence of the effect in temperature dependence of current, when measurements are performed
at constant voltage. Energy gap in active region, as well energy gap in waiveguides change in a nearly 
the same way when temperature changes, but not in exactly the same way. We found such Al concentrations when
the number of QW bound states changes with temperature swap. Figure \ref{temp-swap00} shows how the uppermost 
bound electron state energy, $E_4$ in this case, differs from $E_{CBO}$ 
(energy levels of other QW bound states do not play a significant role in this case), 
for a three Al concentrations in waveguide,
when Al concentration in QW is $8\%$. For Al concentration $34.70 \%$, the $E_4$ energy level exists always 
through temperature swap studied. For Al concentration $34.65 \%$, the $E_4$ energy level does not exist 
between around $150$ and $370 K$. For Al concentration $34.60 \%$, it exists at temperatures higher than
about $470 K$, only. This has profound implications on $I(T)$ dependencies measured at constant voltage 
for these three different Al concentrations of Al in waveguide, as Figure \ref{temp-swap06} illustrates.
For $34.60 \%$ and $34.70 \%$ of Al content, we observe continuous $I(T)$ curves. However, for 
$34.65 \%$ of Al content, at low temperatures $I(T)$ results fall on curve that has been computed for $34.60 \%$
of Al, and at high temperatures they fall on the curve computed for $34.70 \%$ of Al.

%%%%%%%%%%%%%%%%%%%%%%%%%%%%%%%%%%%%%%%%%%%%%%%%%%%%%%%%%%%%%%%%%%%%%%%%%%%%%%%%%%%%%%%%%%%%
\section{Discussion}

In case of tunneling energy barrier, transfer matrix approach is used 
to describe charge transport through it (\cite{Davies}, \cite{Piprek}). 
The interband tunneling current is written as

\begin{equation}\label{current_density}
\begin{array}{ll}
J \sim \int_{E_{min}}^{E_{max}} \cdot N(E) \cdot f(E) \cdot T(E) \cdot dE
\end{array}
\end{equation}

where $T(E)$ is energy-dependent tunneling rate, $N(E)$ is the density of states, 
$f(E)$ is the Fermi-Dirac distribution function, respectively,
and $E_{min}$ and $E_{max}$ are minimum and maximum carrier energies available.

\begin{figure}[h]
\begin{center}
\scalebox{0.9} % Change this value to rescale the drawing.
{
\begin{pspicture}(0,-4.68)(11.561563,4.69)
\definecolor{color39}{rgb}{0.8666666,0.141176470,0.3647}
\definecolor{color39b}{rgb}{0.878431,0.07058823,0.27450}
\definecolor{color187b}{rgb}{0.08235,0.152941,0.996078}
\definecolor{color205b}{rgb}{0.00392,0.898039,0.3411764}
\definecolor{color0b}{rgb}{0.0823529,0.03529,0.19215686}
\definecolor{color0}{rgb}{0.45882352,0.0627450,0.917647}
\definecolor{color22}{rgb}{0.5529411,0.031372,0.376470588}
\psline[linewidth=0.04,linecolor=color0,fillcolor=color0b](4.0215626,2.32)(4.0215626,-3.58)(8.041562,-4.66)(8.061563,1.36)(8.061563,1.36)(8.061563,1.36)
\psline[linewidth=0.04cm,linecolor=color22](0.6015625,4.32)(4.0615625,2.26)
\psline[linewidth=0.04cm,linestyle=dotted,dotsep=0.16cm](1.6015625,-2.4)(9.801562,-2.46)
\psline[linewidth=0.06cm,linestyle=dashed,dash=0.16cm 0.16cm,arrowsize=0.05291667cm 2.0,arrowlength=1.4,arrowinset=0.4]{<->}(5.9415627,1.96)(5.9615626,-2.16)
\psdots[dotsize=0.4,linecolor=color39,fillstyle=solid,fillcolor=color39b,dotstyle=o](3.1215625,3.14)
\psarc[linewidth=0.06,linestyle=dashed,dash=0.16cm 0.16cm,arrowsize=0.05291667cm 2.0,arrowlength=1.4,arrowinset=0.4]{<->}(7.9915624,2.27){1.83}{0.0}{180.0}
\psarc[linewidth=0.06,linestyle=dashed,dash=0.16cm 0.16cm,arrowsize=0.05291667cm 2.0,arrowlength=1.4,arrowinset=0.4]{<->}(4.5615625,3.28){1.38}{0.0}{180.0}
\usefont{T1}{ptm}{m}{n}
\rput(2.00,-2.05){QW bound states}
\usefont{T1}{ptm}{m}{n}
\rput(1.5,1.4){QW continuum states}
\psline[linewidth=0.02cm,arrowsize=0.05291667cm 2.0,arrowlength=1.4,arrowinset=0.4]{<-}(5.9615626,2.62)(3.8615625,2.0)
\psdots[dotsize=0.4,linecolor=color39,fillstyle=solid,fillcolor=color39b,dotstyle=o](9.821563,2.08)
\psdots[dotsize=0.4,linecolor=color39,fillstyle=solid,fillcolor=color187b,dotstyle=o](5.9815626,-2.46)
\psdots[dotsize=0.4,linecolor=color39,fillstyle=solid,fillcolor=color205b,dotstyle=o](6.1215625,2.1)
\psdots[dotsize=0.4,linecolor=color39,fillstyle=solid,fillcolor=color205b,dotstyle=o](5.9415627,3.04)
\psline[linewidth=0.04cm,linecolor=color22](8.081562,1.3)(11.541562,-0.76)
\usefont{T1}{ptm}{m}{n}
\rput(7.8253126,3.6){Thermionic emission}
\usefont{T1}{ptm}{m}{n}
\rput(5.5,-0.21){Carrier Scattering}
\end{pspicture}
}
\caption{The model of carrier scattering at the quantum well used in Sentaurus (based on \cite{tcad}).}
\label{QWscattering}
\end{center}
\end{figure}
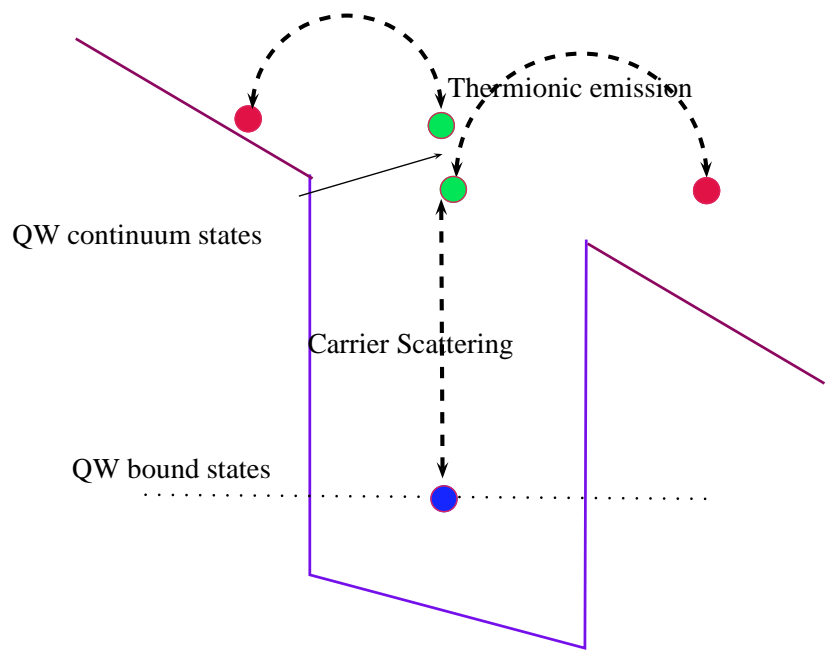

In Sentaurus, the following intuitive model is used to handle the physics of carrier scattering 
at the quantum well (Figure \ref{QWscattering}). The carrier populations are separated into bound and continuum states, 
and separate continuity equations are applied to both populations. The QW scattering model accounts for
the net capture rate, that is, not all of the carriers will be scattered into the bound states of the
quantum well. The electron capture rate from the continuum (subscript $C$) to the bound (subscript $B$) 
states is:

\begin{equation}\label{electron_capture}
\begin{array}{ll}
R = \int _{E_c}^{\infty} dE_C \int _{E_{w}}^{\infty} dE_B \cdot N_C(E_C) \cdot N_B (E_B) \cdot S(E_B,E_C) \cdot f_C (E_C) (1-f_B (E_B))
\end{array}
\end{equation}

where $E_c$ and $E_{w}$ is energy of lowest conduction band-, and bound QW electron states, 
$N(E)$ is the density-of-states, $S(E_B, E_C)$ is the scattering probability, and $f(E)$ is the
Fermi–Dirac distribution.
The reverse process gives the electron emission rate from the bound to continuum states:

\begin{equation}\label{electron_emission}
\begin{array}{ll}
M = \int _{E_c}^{\infty} dE_C \int _{E_{w}}^{\infty} dE_B \cdot N_C(E_C) \cdot N_B (E_B) \cdot S(E_B,E_C) \cdot f_B (E_B) (1-f_C (E_C))
\end{array}
\end{equation}

The net capture rate is $C = R - M$, and for very deep quantum wells (keyword $QWDeep$
must be used for that in Sentaurus) is known to be given by approximation:

\begin{equation}\label{capture_rate_deep}
C = R - M = \left(1 - exp(\eta_B - \eta_C)\right) \cdot \frac {n_C}{\tau}
\end{equation}

where $\eta_B = (-q \Phi_B - E_C)/k_B T$ and $\eta_C = ( -q \Phi_C - E_C)/k_B T$
contain the quasi-Fermi level information and $\tau$ is the capture time. 
The capture time represents scattering processes attributed to carrier–carrier and carrier-LO phonon 
interactions involving bound quantum well states. The net capture rate $C$ 
is added to the continuity equations as a recombination term. 

In a similar way scattering of holes is computed, with their own characteristic 
capture time. These parameters are specified in Sentaurus by the keywords $QWeScatTime$ and
$QWhScatTime$. Their default values, $8 \cdot 10^{-13} s$ and $4 \cdot 10^{-13} s$, respectively, 
correspond reasonably well to these based on theory (\cite{Hernandez}, \cite{Blom}). 
Photoluminescence spectroscopy results give values of an order of $3-20 ps$ \cite{Blom}.
In most of our modelling, if not indicated otherwise, we use also default values
of electron and hole mobility, represented in Sentaurus by parameters
$eQWMobility = 9200 cm^2/Vs$ and $hQWMobility = 400 cm^2/Vs$.

For shallow quantum wells, the energy transfer during scattering can only occur in a limited
range. In the limit of elastic scattering, the net capture rate is then approximated by:

\begin{equation}\label{capture_rate_shallow}
C = \left(\frac{F_{3/2}(\eta_C)}{F_{1/2}(\eta_C)} -  \frac{F_{3/2}(\eta_B)}{F_{1/2}(\eta_B)}\right) \cdot \frac {n_C}{\tau}
\end{equation}

where $F_m$ is the Complete Fermi-Dirac integral of the order of $m$. The shallow quantum well model is
activated by the keyword $QWShallow$. 

It should be pointed out that Equations \ref{capture_rate_deep} and \ref{capture_rate_shallow}, for deep
and shallow quantum wells, respectively, while provide a convenient,
intuitive description of carriers scattering and capture on QW, these are approximate only.
In particular, there is no dependence of capture time on energy of unbound carriers there 
and no periodic oscillations as a function of QW size (\cite{Hernandez}, \cite{Blom}). 
GaAs/AlGaAs are considered to have deep quantum wells. However, as we have shown, the results
of our modelling indicate on a strong role of bound QW states located very closely to the 
offset energy levels of quantum wells. For these reasons, we did not restrict our calculations to 
deep- or shallow- QW models but used instead the full model available in Sentaurus.

Equations \ref{current_density} - \ref{electron_emission} all depend on density of bound states in QW. 
We expect hence that current through the QW will be proportional to the density of all bound states in
QW. In effective mass approximation, the two-dimensional density of electron states
within each QW subband $n$ equals (\cite{Piprek})

\begin{equation}\label{density_of_states}
\begin{array}{ll}
N_n(E) = \frac{m_n}{\pi \hbar ^2},~~~ for~ E > E_n
\end{array}
\end{equation}

Hence, the current should be proportional to the \emph{number of bound states} $\times$ \emph{carrier mass}.
The quantity computed this way (with a certain multiplication factor) 
is represented by large circles in Figure \ref{qw300_talk0A}. 
Though it must not be exact (for instance, no difference
in scattering rates for electrons and holes is accounted for), it fits reasonably well $I(d_a)$ dependence.

%%%%%%%%%%%%%%%%%%%%%%%%%%%%%%%%%%%%%%%%%%%%%%%%%%%%%%%%%%%%%%%%%%%%%%%%%%%%%%%%%%%%%%%%%%%%%%%%%

\section{Summary and Conclusion}

When performing modeling of laser characteristics as a function of the width of active region 
we noticed a non-monotonic, discontinuous dependence of $I(d_a)$ (when measured at constant voltage applied). 
A careful analysis of the data led us to the hypothesis that discontinuities occur when the most 
upper QW, bound energy states are found very close to the conduction or valence band energy offsets. 
The effect, hence, is thought to be related to changes in density of states of carriers from one hand,
and to fast changes in carrier transfer matrix through QW for QW bound states close 
to $E_{CBO}$ or $E_{VBO}$. As such, it ought to be more pronounced at lower temperatures. 
Indeed, results of modeling $I(d_a)$ at liquid Nitrogen temperature (77.6 K) confirmed this idea. 

The effect is observed also when modelling current as a function of QW depth (Al concentration in waveguide).

Therefore, we concluded that a similar effect will be present also in modelling $I$ as a function
of temperature. In that case however a carefull design of laser properties is needed, in such a way
that a transition of the most upper QW energy state will pass through an endge of quantum well when temperature is swapt. 

These observations are important for proper designing of semiconducting lasers 
(choice of Al concentrations, thickness of the active region, etc). However, they also illustrate well
the intrinsic carrier transport mechanisms in SCH lasers. Potentially, might be useful as a kind
of quantum level spectroscopy tool when testing laser designs for technological applications.

%%%%%%%%%%%%%%%%%%%%%%%%%%%%%%%%%%%%%%%%%%%%%%%%%%%%%%%%%%%%%%%%%%%%%%%%%%%%%%%%%%%%%%%%%%%%%%%%%

%%%%%%%%%%%%%%%%%%%%%%%%%%%%%%%%%%%%%%%%%%%%%%%%%%%%%%%%%%%%%%%%%%%%%%%%%%%%%%%%%%%%%%%%%
\clearpage

%%%%%%%%%%%%%%%%%%%%%%%%%%%%%%%%%%%%%%%%%%%%%%%%%%%%%%%%%%%%%%%%%%%%%%%%%%%%%%%%%%%%%%%%%%%%%
%----------------------------------------------------
% Generated with LaTeXDraw 2.0.8
% Fri Nov 04 18:14:29 MSK 2011
% \usepackage[usenames,dvipsnames]{pstricks}
% \usepackage{epsfig}
% \usepackage{pst-grad} % For gradients
% \usepackage{pst-plot} % For axes
%%%%%%%%%%%%%%%%%%%%%%%%%%%%%%%%%%%%%%%%%%%%%%%%%%%%%%%%%%%%%%%%%%%%%%%%%%%%%%%%%%%%%%%%%%%%%

\end{document}